\documentclass{pnastwo}

\usepackage{graphicx}
\usepackage{amssymb,amsfonts,amsmath}
\usepackage{color} 

\newcommand{\bp}{{\bf p}}
\newcommand{\glog}{\Lambda}
\newcommand{\gexp}{{\cal E}}

\newcommand{\mystar}{*}

\newcommand{\gL}{{\cal L}}
\newcommand{\gG}{{\cal G}}

\newcommand{\SI}{supporting information}

\begin{document}


\title{Generalized entropies and logarithms and their duality relations 
}

\author{Rudolf Hanel\affil{1}{Section for Science of Complex Systems, Medical University of Vienna, Spitalgasse 23, 1090 Vienna, Austria}, Stefan Thurner\thanks{To whom correspondence should be addressed. E-mail:
stefan.thurner@meduniwien.ac.at}\affil{1}{}\affil{2}{Santa Fe Institute, 1399 Hyde Park Road, Santa Fe, NM 87501, USA}
\and Murray Gell-Mann\affil{2}{}}


\maketitle


\begin{article}

\begin{abstract} 
For statistical systems that violate one of the four Shannon-Khinchin axioms, entropy 
takes a more general form than the Boltzmann-Gibbs entropy. The framework of superstatistics allows one to formulate 
a maximum entropy principle with these generalized entropies, making them useful for understanding distribution functions 
of non-Markovian or non-ergodic complex systems. For such systems where the composability axiom is violated there exist 
only two ways to implement the maximum  entropy principle, one using  escort probabilities, the other not. The two ways are connected through a duality. 
Here we show that this duality fixes a unique escort probability, which allows us to derive a complete theory of the generalized logarithms
that naturally arise from the violation of this axiom. We then show how the functional forms of these generalized logarithms 
are related to the asymptotic scaling behavior of the entropy. 
\end{abstract}


\keywords{thermodynamics | entropy | classical statistical mechanics | correlated systems}

\dropcap{T}he
concept of {\em superstatistics} \cite{beck03,beck05,beckEPL03} provides a formal framework 
for a  wide class of generalizations of statistical mechanics that were introduced recently. 
Within this framework it is possible to formulate a maximum entropy principle, even for 
non-ergodic or non-Markovian systems, including many complex systems.  
From an axiomatic point of view, non-additive systems are characterized by the fact that the fourth 
Shannon-Khinchin (SK) axiom\footnote{
	Shannon-Khinchin axioms:
	(i) Entropy is a continuous function of the probabilities $p_i$ only, i.e. $s$ should not explicitly depend on 
	any other parameters. 
	(ii) Entropy is maximal for the equi-distribution $p_i=1/W$.
	-- From this the concavity of $s$ follows. 
	(iii) Adding a state $W+1$ to a system with $p_{W+1}=0$ does not change the entropy of the system. 
	-- From this $s(0)=0$ follows.
	(iv) Entropy of a system composed of 2 sub-systems $A$ and $B$, is $S(A+B)=S(A)+S(B |A)$.
	\label{foo_shannon}
},
governing composability of statistical systems, is violated. 
For systems where all four SK axioms hold, the entropy is uniquely determined as the  
Boltzmann-Gibbs-Shannon entropy \cite{shannon,Khinchin}, $S_{BGS}=-k\sum p_i \log p_i$.
In the case where only the first three axioms are valid (e.g. non-Markovian systems) 
the entropy has a more general form \cite{HTclassification}.
In the thermodynamic limit -- which captures the asymptotic behavior for small values of the $p_i$ --
the entropy is given by the formula 
\begin{equation}
S_{cd} \propto \sum_i \Gamma(1+d,1 -c \log p_i) \,, 
\label{ent}
\end{equation}
where $\Gamma$ is the incomplete gamma function and 
$(c,d)$ are constants that are uniquely determined by the scaling properties of the statistical system in its thermodynamic limit. 
In previous work \cite{HTG} we were able to show that for systems where the first three SK axioms hold, 
there exist only two ways to formulate a consistent maximum entropy principle. 
Starting with an entropy of  ``trace form'' 
\begin{equation}
	S[\bp]=\sum_{i=1}^W s(p_i) \,, 
\label{ents}
\end{equation} 
the maximization condition becomes $\delta \Phi=0$ with
\begin{equation}
	\Phi[p]=S[p]-\alpha\left(\sum_i p_i-1\right)-\beta \left(\sum_i  Q_i[p,s] \epsilon_i -U \right)\,,
	\label{MEP}
\end{equation}
where the last two terms are the constraints. 
The first of the two possible approaches ($HT$ approach) \cite{hanel07, thurner08}, 
uses a generalized entropy and the usual form of the 
constraint, $Q_i^{HT}[p]= p_i$. 
The other approach ($TS$ approach), suggested in \cite{TsallisSouza}, uses a generalized 
entropy {\em and} a more general way to impose  constraints  
\begin{equation}
	Q_i^{TS}[p,s]=P_i[p,s] = \frac{p_i+\nu s(p_i)}{\sum_j p_j+\nu s(p_j)}\, .
	\label{escort-distr}
\end{equation}
$P_i$ is a so-called {\it escort probability} and $\nu$ is a real number. 
While in the HT case the constraint has the usual interpretation as an energy constraint, 
we do not attempt to give a physical interpretation of the escort probabilities.
The two approaches have been shown to be connected by a duality map 
$\mystar:S_{HT} \stackrel{ \mystar}{ \leftrightarrow} S_{TS}$, with 
$\mystar\mystar$ 
(meaning applying * twice) 
being the identity \cite{HTG}. A special case of this duality has been observed in \cite{Ferri2005}.

Entropies can be conveniently formulated using their associated generalized logarithms. 
We first specify the space $\gL$ of {\em proper generalized logarithms} $\glog\in\gL$. 
We  consider a generalized logarithm to be {\em proper} if the following properties hold: 
\begin{itemize}
\item $\glog$ is a differentiable function $\glog:\mathbb{R}_+\to\mathbb{R}$.
This is necessary for a finite second derivative of the entropy.  
\item $\glog$ is monotonically increasing, which is a consequence of the second SK axiom.
\item $\glog(1)=0$, captures the requirement that the entropy of single-state systems is $0$. 
\item $\glog'(1)=1$, is needed to fix the units of  entropy. 
 
\end{itemize}
In both approaches ($HT$ and $TS$) there exist proper generalized logarithms
$\glog_{HT}$ and $\glog_{TS}$ such that 
\begin{equation}
s_{HT}(p_i)=-k\int_0^{p_i} dx\,\glog_{HT}(x/x_0)\,,
\label{sht}
\end{equation}
and
\begin{equation}
s_{TS,\nu}(p_i)=-k\int_0^{p_i} dx\,\glog_{TS,\nu}(x/x_0)\, , 
\end{equation}
with $x_0$ a constant. If both approaches predict the same distribution function $\bp=\{p_i\}_{i=1}^W$ as a result of the maximization 
of Eq. (\ref{MEP}),  then it can be shown that
the two entropic functions $s_{HT}$ and $s_{TS}$ are one-to-one related by 
\begin{equation}
\frac{1}{\glog_{TS,\nu}(x)}-\frac{1}{\glog_{HT}(x)}=k\nu\,.
\label{HTTSduality}
\end{equation}
In the following we set $k=1$. This can be achieved either by choosing physical units accordingly, or by 
simply absorbing $k$ into $\nu$, so that $\nu$ becomes a dimensionless parameter.

The full implications of Eq. (\ref{HTTSduality}), which is related to the essence of this paper, can be summarized as follows. 
The statistical properties of a physical system, for instance a superstatistical system as discussed in \cite{HTG}, 
uniquely determine the entropy $S_{HT}$. A priori, there exists a spectrum of $TS$-entropies, $S_{TS,\nu}$,
whose boundaries are determined by the properties of the generalized logarithm associated with $S_{HT}$. 
Moreover, these properties determine a particular value $\nu^*$, so that $S_{TS,\nu^*}$ and $S_{HT}$ 
become a pair of dual entropies.
This unique duality allows us to derive a complete theory of generalized logarithms naturally arising as a 
consequence of the fourth SK axiom being violated.
We present a full understanding of how the $TS$ and the $HT$ approaches are interrelated
and derive the most general form of families of generalized logarithms that are compatible
with a maximum entropy principle and the first three SK axioms. Finally, we
demonstrate how these logarithms can be classified according to their asymptotic scaling properties, following the 
results presented in \cite{HTclassification}.

\section{The duality}

In contrast to the images of generalized logarithms, which need not span $\mathbb{R}$ completely and can differ from one another, the domain of generalized logarithms is always all of $\mathbb{R}_+$.
For these reasons one may classify generalized logarithms according to the minimum and maximum values of their images and consider the group $G$ of order-preserving automorphisms on $\mathbb{R}_+$ that keep an infinitesimal neighborhood of $1\in\mathbb{R}_+$ invariant, 
as the means to generate these classes. 
In the following we call the elements $g$ 
of this automorphism group {\it scale transformations}. More precisely,
$g\in\gG$ is a  scale transformation if $g$ is differentiable and maps $\mathbb{R}_+$  to $\mathbb{R}_+$ one-to-one, 
$g'>0$, 
$g(1)=1$, and 
$g'(1)=1$.
From these properties it follows that $g(0)=0$ and
$\lim_{x\to\infty} g(x)=\infty$.
Finally, we use the notation $f\circ g(x)=f(g(x))$. 
  
Scale transformations leave the image of a generalized logarithm invariant. This allows us
to parametrize classes in the following way.
Given a proper generalized logarithm $\glog\in\gL$, we write for its maximum and minimum values
\begin{equation}
\underline{\glog} \equiv \min\{\glog(x)|x\in\mathbb{R}^+\}\quad,\quad 
\overline{\glog} \equiv \max\{\glog(x)|x\in\mathbb{R}^+\}, , 
\end{equation}
and define  two functionals
\begin{equation}
\nu_-[\glog] \equiv -\frac{1}{\overline{\glog}}\leq0 \quad ,\quad  \nu_+[\glog] \equiv -\frac{1}{\underline{\glog}}\geq0 \, , 
\label{nunu_main}
\end{equation}
which associate numbers $\nu_+$ and $\nu_-$ to any $\glog$. 
For their sum we write $\nu^*=\nu_++\nu_-$.
Next, we define sets of proper generalized logarithms, 
\begin{equation}
\gL_{\nu_+,\nu_-}=\left\{\glog\in\gL\,|\, \nu_+=\nu_+[\glog]\,\,{\rm and}\,\,\nu_-=\nu_-[\glog]\right\}\, .
\label{maxminset}
\end{equation}
Members of $\gL_{\nu_+,\nu_-}$ all have the same maximum and minimum values. 
In fact, the $\gL_{\nu_+,\nu_-}$ are exactly the equivalence classes in $\gL$ generated by $G$:
Two generalized logarithms $\glog^{(A)}$ and $\glog^{(B)}$ are considered equivalent if there exists a scale transformation $g\in G$ such that $\glog^{(B)}=\glog^{(A)}\circ g$.  
The space of generalized logarithms can be written as the union of these sets, $\gL= \bigcup_{\nu_+,\nu_-} \gL_{\nu_+,\nu_-}$. 

With these definitions we now analyze the relation between the $HT$ and $TS$ approaches. 
Assuming that $\glog_{HT}$ is given, Eq. (\ref{HTTSduality}) implies    
\begin{equation}  
\glog_{TS,\nu}(x)=T_\nu\circ \glog_{HT}(x)\,,\quad T_\nu(x)=\frac{1}{\frac1x+\nu} \, .
\end{equation}  
$T_\nu$ is a shift operator with the property $T_\nu \circ T_\mu=T_{\nu+\mu}$.  
We have of course $\glog_{TS,0}=\glog_{HT}$. 
The fact that $\glog_{HT}$ is a proper generalized logarithm does not imply that $\glog_{TS}$ is also proper for all choices of $\nu$. 

In fact, given that $\glog\in\gL$, it can be shown (see \SI) that $T_\nu\circ\glog \in \gL$ if and only if 
$\nu_-[\glog]\leq\nu\leq\nu_+[\glog]$. 
Moreover, for $\glog\in\gL_{\nu_+,\nu_-}$ and for $T_\nu \circ \glog$ being a proper generalized logarithm it follows that 
$T_\nu \circ \glog\in \gL_{\nu_+-\nu,\nu_--\nu}$.
As a consequence $\glog_{TS,\nu}(x)=T_\nu\circ \glog_{HT}(x)$ is  proper only for 
$\nu_-[\glog_{HT}] \leq\nu\leq\nu_+[\glog_{HT}]$, and 
\begin{equation} 
\glog_{TS,\nu}\in\gL_{\nu_+-\nu,\nu_--\nu}\Leftrightarrow \glog_{HT}\in\gL_{\nu_+,\nu_-} \, .
\end{equation}

This equation does not uniquely determine a duality relation $\mystar$ on $\gL$, 
yet by imposing the condition that $\mystar$ commute with scale transformations $g\in\gG$, 
it can be shown (see \SI) that $\mystar$
is given by
\begin{equation}
\glog^{\mystar}=T_{\nu_++\nu_-}\circ\glog\,
\quad\mbox{for}\quad\glog\in\gL_{\nu_+,\nu_-}\, ,
\label{theor2equ}
\end{equation}
with the property
\begin{equation}
\glog\in\gL_{\nu_+,\nu_-}\Leftrightarrow \, \glog^{\mystar}\in\gL_{-\nu_-,-\nu_+}\,.
\label{dualsets}
\end{equation}

Thus for each $\glog_{HT}$ there exists a unique value 
$\nu^*=\nu_+[\glog_{HT}]+\nu_-[\glog_{HT}]$ 
such that $\glog_{TS,\nu^*}$ 
is a proper generalized logarithm.
The duality map $*$ gives $\glog_{TS,\nu^*}=\glog_{HT}^\mystar$.
Furthermore, since $\mystar$ and $g$ commute 
($(\glog\circ g)^\mystar=\glog^\mystar\circ g$), {\em any} proper generalized logarithm $\glog$  
can be decomposed into a specific representative $\glog_{\nu_+,\nu_-}\in\gL_{\nu_+,\nu_-}$,
and a scale transformation $g$, so that 
\begin{equation}
\glog=\glog_{\nu_+,\nu_-}\circ g \,. 
\label{prop}
\end{equation} 
This implies that {\em any}
$\glog_{HT}$ or $\glog_{TS,\nu}$ can be decomposed in this way and that the dual logarithms, 
$\glog_{HT}$ and $\glog_{HT}^*=\glog_{TS,\nu^*}$ transform
identically under scale transformations.

\section{The functional form of the generalized logarithms}

Equation (\ref{dualsets}) implies the existence of transformations 
that map members of $\gL_{\nu_+,\nu_-}$ to members of  
$\gL_{-\nu_-,-\nu_+}$. These maps can be used to represent the duality $*$ 
on specific families $\glog_{\nu_+,\nu_-}\in\gL_{\nu_+,\nu_-}$.
$\glog(x)\to-\glog(1/x)$ is exactly such a map, since 
$\max\{-\glog(1/x)\,|\,x\in\mathbb{R}_+\}=\max\{-\glog(x)\,|\,x\in\mathbb{R}_+\}=-\min\{\glog(x)\,|\,x\in\mathbb{R}_+\}=-\underline{\glog}$.
The same holds for $\min\{-\glog(1/x)\,|\,x\in\mathbb{R}_+\}=-\overline{\glog}$.
This  allows us to construct $\glog_{\nu_+,\nu_-}$ with the properties
\begin{equation}
	\glog^\mystar_{\nu_+,\nu_-}(x)=\glog_{-\nu_-,-\nu_+}(x)=-\glog_{\nu_+,\nu_-}(1/x) \,.
	\label{central-prop}
\end{equation}
By using Eq. (\ref{theor2equ}) and inserting $\glog^\mystar_{\nu_+,\nu_-}(x)=-\glog_{\nu_+,\nu_-}(1/x)$ 
into Eq. (\ref{HTTSduality}), we get
\begin{equation}
	\frac{1}{\glog_{\nu_+,\nu_-}(1/x)}+\frac{1}{\glog_{\nu_+,\nu_-}(x)}=-(\nu_++\nu_-) =-\nu^*  \,.
	\label{HTTSduality2}
\end{equation}
This equation may have many solutions $\glog_{\nu_+,\nu_-}$, but we can restrict ourselves to 
finding a particular one. All the others can be obtained by scale transformations. This is seen as follows:  
Suppose $\glog_{\nu_+,\nu_-}^{(A)}$ and  $\glog_{\nu_+,\nu_-}^{(B)}$ are both solutions of Eq. (\ref{HTTSduality2}); 
then according to  Eq. (\ref{prop}) for any pair $(\nu_+,\nu_-)$ there exists a scale transformation $\tilde g_{\nu_+,\nu_-}$ 
such that $\glog_{\nu_+,\nu_-}^{(B)} = \glog_{\nu_+,\nu_-}^{(A)}  \circ \tilde g_{\nu_+,\nu_-}$.
Since $\tilde g_{\nu_+,\nu_-}$ must leave Eq. (\ref{central-prop}) 
invariant 
(this is not the case for arbitrary scale transformations $g\in \gG$), 
these scale transformations have two properties. The first is $\tilde g_{\nu_+,\nu_-}(x) \tilde g_{\nu_+,\nu_-}(1/x)=1$, 
which makes them members of a subgroup 
$\tilde g\in \gG_0\subset \gG$ of all possible scale transformations $g\in\gG$. 
The second property is $\tilde g_{\nu_+,\nu_-}=\tilde g_{-\nu_-,-\nu_+}$ and follows from the fact that $*$ 
commutes with scale transformations.

A particular solution of Eq. (\ref{HTTSduality2}) is given by
\begin{equation}
\glog_{\nu_+,\nu_-}(x)=\left(\frac{1}{\frac{2}{\nu_+-\nu_-}h\left(\frac{\nu_+-\nu_-}{2}\log(x)\right)}-\frac{\nu_++\nu_-}{2} \right)^{-1}\,, 
\label{family1}
\end{equation}
with $h:\mathbb{R}\to[-1,1]$  a continuous, monotonically increasing, odd function, 
with $\lim_{x\to\infty}h(x)=1$ and $h'(0)=1$.
It can easily be verified that this solution has all the required properties: 
$\glog_{\nu_+,\nu_-}$ is a proper logarithm with $\glog_{\nu_+,\nu_-}\in\gL_{\nu_+,\nu_-}$ (correct minimum and maximum),
$\glog^*_{\nu_+,\nu_-}(x)=-\glog_{\nu_+,\nu_-}(1/x)=\glog_{-\nu_-,-\nu_+}(x)$, $\glog_{\nu,-\nu}(x)=h(\nu \log(x))/ \nu$ is self-dual, 
and  $\lim_{\nu_+\to0}\lim_{\nu_-\to0}\glog_{\nu_+,\nu_-}(x)=\log(x)$.

This means that we can generate a specific family of logarithms $\glog_{\nu_+,\nu_-}$, following 
Eq. (\ref{central-prop}),
by choosing one particular function $h$ (for instance $h(x)=\tanh (x)$) and then using scale transformations 
to reach all other possibilities. In particular, some family
$\tilde\glog_{\nu_+,\nu_-}$ with the property $\tilde\glog^*_{\nu_+,\nu_-}(x)=\tilde\glog_{-\nu_-,-\nu_+}(x)$ 
can be reached by a family of scale transformations
$\tilde g_{\nu_+,\nu_-}=\gexp_{\nu_+,\nu_-}\circ \tilde\glog_{\nu_+,\nu_-}\in \gG$, where $\gexp_{\nu_+,\nu_-}\equiv \glog^{-1}_{\nu_+,\nu_-}$
are generalized exponential functions (inverse functions of logarithms). Moreover, if
$\tilde\glog_{\nu_+,\nu_-}$ also follows Eq. (\ref{central-prop}), then $\tilde g_{\nu_+,\nu_-}\in \gG_0$.

The family of dual logarithms discussed in \cite{HTG} is obtained in the framework presented here by
setting either $\nu_+=0$ or $\nu_-=0$. 
These classes correspond to logarithms that are unbounded either from below or from above while
the duality maps $\gL_{\nu,0} \stackrel{ \mystar}{ \leftrightarrow} \gL_{0,-\nu}$. 
Moreover, in \cite{HTG} only pairs of dual logarithms 
have been considered such that $\glog^*(x)= -\glog(1/x)$, and the part scale transformations play 
in the {\it unique} definition of $\mystar$ had not yet been described. 

We are now in a position to understand all observable distribution functions emerging from the two approaches
in terms of a single two-parameter family of generalized logarithms $\glog_{\nu_+,\nu_-}$ and a scale transformation. 
This result now raises the question of how $\glog_{\nu_+,\nu_-}$ is related to the two-parameter logarithms 
associated with the $(c,d)$-entropies in Eq. (\ref{ent}),  \cite{HTclassification}. 
That will further clarify the role of the scale transformations. 

\section{The $\glog_{\nu_+,\nu_-}$ logarithm and $(c,d)$-entropy}

Generalized entropies can be classified with respect to their asymptotic scaling behavior in 
terms of two scaling exponents $c$ and $d$,   
where $0<c\leq1$ and $d$ is a real number \cite{HTclassification}.
They are obtained from the scaling relations 
\begin{equation}
\lambda^{c}=\lim_{x \to 0 }\frac{s(\lambda x)}{s(x)} \quad , \quad (1+a)^d=\lim_{x \to 0 }\frac{s(x^{1+a})}{x^{ac}s(x) } \, ,
\label{scale}
\end{equation}
where $s$ is the summand in Eq. (\ref{ents}). Using Eqs. (\ref{scale}), de l'H\^opital's rule, and  the fact that 
$s'(x)=-\glog(x)$, we find the exponents $(c,d)$ for a given $\glog\in\gL$ 
\begin{equation}
\lambda^{c-1}=\lim_{x \to 0 }\frac{\glog(\lambda x)}{\glog(x)}\, ,\quad 
(1+a)^d=\quad \lim_{x \to 0 }\frac{\glog(x^{1+a})}{x^{a(c-1)}\glog(x) } \,,
\end{equation}
where we represent $\glog$ as $\glog=\glog_{\nu_+,\nu_-}\circ g$. 
In this way we get the dependence of $(c,d)$ as a function of $(\nu_+,\nu_-)$, $h$, and the scale transformation $g$. 
We first compute the asymptotic properties of $h$ and $g$, defining the exponents $c_{h,g}$ and $d_{h,g}$ by   
\begin{equation}
\lambda^{c_{h,g}}=\lim_{x\to 0}\frac{\varphi_{h,g}(\lambda x)}{\varphi_{h,g}(x)}\,,
\quad (1+a)^{d_{h,g}}=\lim_{x\to 0}\frac{\varphi_{h,g}(x^{1+a})}{x^{a c_{h,g}}\varphi_{h,g}(x)}\,, 
\label{ghprops}
\end{equation}
where  $\varphi_{h,g}=1+h\circ\log\circ g(x)$. Note that $\log\circ g\in\gL_{0,0}$. By  defining $\glog_0 \equiv \log\circ g$ we
compute its scaling exponents $c_0$ and $d_0$ 
\begin{equation}
\lambda^{c_0-1}=\lim_{x\to 0}\frac{\glog_0(\lambda x)}{\glog_0(x)}\,,
\quad (1+a)^{d_0}=\lim_{x\to 0}\frac{\glog_0(x^{1+a})}{x^{a (c_0-1)}\glog_0(x)}\,.
\label{gprops}
\end{equation}
With these preparations one can derive the results  
\begin{eqnarray}
c&=&\left\{\begin{array}{ll} 
1 & {\rm for }\,\,\nu_+\neq 0 \,,\\ 
1-c_{h,g} &  {\rm for }\,\,\nu_+= 0 \,\, {\rm and} \,\, c_0\neq 1 \,,\\
1-c_{h,g}\left(-\frac{\nu_-}{2}\right)^{\frac{1}{d_0}} &  {\rm for }\,\,\nu_+= 0 \,\, {\rm and} \,\, c_0=1 \,,
\end{array}\right. 
\nonumber \\
d&=&\left\{ \begin{array}{ll} 
0 & {\rm for }\,\,\nu_+\neq 0 \,, \\ 
-d_{h,g} &  {\rm for }\,\,\nu_+= 0  \,.
\end{array}\right.
\label{dprop}
\end{eqnarray}
This  demonstrates clearly that, given a fixed  $h$, $c$ is controlled by $\nu_-$, (for $\nu_+=0$ and $c_0=1$), 
while $d$ is determined by the scale transformation. 

\section{Examples}

{\bf Example 1 - A simple choice for $h$}: 
For example, fix  $h(x)= \tanh(x)$. From Eq. (\ref{family1}) we get for the generalized logarithm 
\begin{equation}
	\glog_{\nu_+,\nu_-}(x) = \frac{x^{\nu_+-\nu_-}-1}{\nu_+-\nu_-x^{\nu_+-\nu_-}} \, .
	\label{ex1}
\end{equation}
The associated generalized exponential (inverse of the generalized logarithm) is 
\begin{equation}
	{\cal E}_{\nu_+,\nu_-}(x)  = \left( \frac{1+ \nu_+ x}{1+\nu_-x} \right)^{\frac{1}{\nu_+-\nu_-}} \, .
	\end{equation}
\linebreak
{\bf Example 2 - Power laws}: 
By setting $h(x)= \tanh(x)$ and $\nu_+=0$, we get from Eq. (\ref{ex1})  the so-called $q$-logarithm, 
with $\glog_{0,\nu_-}=\log_q(x)\equiv (1-(1-q)x)^{1/(1-q)}$, 
where $0\leq q = 1-\nu_- \leq 1$.
The dual is $\glog_{0,\nu_-}^{\mystar}(x)=\frac{x^{-\nu_-}-1}{-\nu_-}=\log_{2-q}(x)$, 
and we recover the well known duality for $q$-logarithms. 
It is also well known that  $\log_q$ results from the use of escort distributions \cite{TMP,tsallisbook,TsallisSouza}, 
while  $\log_{2-q}$ is a natural result of the $HT$ approach \cite{hanel07,thurner08}.  

An example of a generalized logarithm that is not a power is obtained by taking $\nu_- = - \frac{\nu_+}{2}$  in  Eq. (\ref{ex1}).
One obtains $\glog_{\nu_+, - \frac{\nu_+}{2}}=  (x^{-\frac{3}{2}\nu_+}-1)/(\nu_+ (1+\frac12 x^{\frac32 \nu_+} ))$, with the dual 
$\glog_{\nu_+, - \frac{\nu_+}{2}} ^{\mystar} =  (x^{\frac{3}{2}\nu_+}-1)/(\nu_+ (\frac12+ x^{\frac32 \nu_+} ))$.
\\
\linebreak
{\bf Example 3 - Scale transformations}: Any proper generalized logarithm can be written 
as a composition of a representative logarithm from Eq. (\ref{family1}) and a scale transformation,
with $\glog = \glog_{\nu_+ , \nu_-} \circ g$.
For example pick $\glog_{0 ,0}(x)=\log(x)$, and $g_d(x)= \exp [1-(1- d\log(x) )^{\frac1d} ]$, where $d>0$ is a parameter of $g$. 
The generalized logarithm then becomes 
\begin{equation}
\glog (x) = \glog_{0 ,0}( g(x) ) = 1-\left(1- d\log(x) \right)^{\frac1d} \, .
\end{equation}
The associated generalized exponential is a stretched exponential, ${\cal E}_d(x)=\exp(-\frac1d[(1-x)^d-1 ])$, which is
the  known result for $(c,d)$-entropies with $c=1$ and $d>0$, \cite{HTclassification,HTextensive}. 
\\
\linebreak
{\bf Example 4 - Different choices for $h$}:
Suppose that a physical situation demands a specific $\glog$ and two observers $A$ and $B$ choose to represent $\glog$ differently. 
Observer $A$ chooses $h_A(x)= \frac{2}{\pi} \arctan(\frac{\pi}{2}x)$ to represent $\glog$, 
so that $\glog = \glog^{(A)}_{\nu_+,\nu_-} \circ \tilde g^{(A)}$, 
and observer $B$ chooses $h_B(x)=\tanh (x)$ to represent $\glog = \glog^{(B)}_{\nu_+,\nu_-} \circ \tilde g^{(B)}$. 
Then $\glog^{(A)}_{\nu_+,\nu_-}$ and $\glog^{(B)}_{\nu_+,\nu_-}$ can only differ by a
scale transformation $\tilde  g_{\bar \nu} \in \gG_0$ with $\bar \nu \equiv \frac{\nu_{+}-\nu_-}{2}$ and it follows 
that $\tilde g_{\bar \nu} = \exp [ \frac{2}{\bar\nu} h^{-1}_{A} \circ h_{B} (\frac{\bar\nu}{2} \log (x) )  ]$. 
For the particular functions $h_A$ and $h_B$ we have chosen, we get
$\tilde g_{\bar \nu} (x) = \exp [\frac{4}{\bar \nu \pi} \tan(\frac{\pi}{2} \frac{x^{\bar \nu}-1}{x^{ \bar \nu} + 1})]$.

\section{Discussion}
By studying the two types of entropies that are related to the two possible ways to formulate a maximum entropy 
principle for systems that explicitly violate the fourth SK axiom, we find that there exists a {\em unique} duality that relates 
the two entropies. 
Consequently thermodynamic properties derived from those two entropies will also be related through the duality.
We show that the maximum and minimum of $\glog_{HT}$  determine a unique value $\nu^*$ for which 
$\glog_{TS,\nu^*}$ is the dual of $\glog_{HT}$. In this way it is possible for an object such as $\glog_{HT}$,
which does not explicitly carry an index $\nu$, to become dual to an object that does, such as $\glog_{TS,\nu}$.  
The existence of this duality opens the way to characterizing all possible generalized logarithms as compositions of a specific functional form $\glog_{\nu_+,\nu_-}$ and scale transformations $g$. 
We derive the explicit form of  $\glog_{\nu_+,\nu_-}$ and show that these logarithms are one-to-one related to two 
asymptotic scaling exponents $(c,d)$ which allow one to characterize strongly 
non-ergodic or non-Markovian systems in their thermodynamic limit \cite{HTclassification}. 
$\nu_-$ is shown to be directly related to $c$, 
while the form of the scale transformation $g$  determines  $d$.
In summary, we provide a complete theory of all generalized logarithms that can arise as a consequence of the violation of the fourth SK axiom.

\begin{acknowledgments}
This material is based upon work supported in part by the National Science Foundation, grant \# 1066293
and the hospitality of the Aspen Center for Physics.

R.H. and S.T. thank the SFI for hospitality. M. G.-M. is glad to acknowledge the 
generous support of Insight Venture Partners and the Bryan J. and June B. Zwan Foundation.
\end{acknowledgments}

\newpage 

\noindent
{\large \bf Supporting Information}
\\

%
{\bf Theorem 1} Let $\glog\in\gL$. Then $T_\nu \circ \glog\in\gL$ if and only if  
$\nu_-[\glog]\leq\nu\leq\nu_+[\glog]$. 
Moreover, if $\glog\in\gL_{\nu_+,\nu_-}$ and $T_\nu \circ \glog$ is a proper generalized 
logarithm, then 
$T_\nu \circ \glog\in \gL_{\nu_+-\nu,\nu_--\nu}$.
\linebreak
\linebreak
{\bf Proof of Theorem 1.}
We recall that $T_\nu$ is defined as $T_\nu(x)=(1/x+\nu)^{-1}$. 
If there exists an $x>0$ such that $\glog(x)=-1/\nu$, then $T_\nu \circ\glog(x)$ possesses 
a pole where $T_\nu \circ\glog(x)$ changes sign. 
In that case $T_\nu\circ\glog$ is neither a continuous nor a monotonically increasing 
function and therefore not a proper generalized logarithm. 
Conversely, if no $x>0$ exists such that $\glog(x)=-1/\nu$, then $T_\nu \circ\glog$ has no pole
and is a continuous monotonically increasing function since $T'_\nu(x)=(1+\nu x)^{-2}>0$ for all $x$.
Moreover, $T_\nu\circ \glog(1)=0$ and $(T_\nu\circ \glog)'(1)=T'_\nu(0)\glog'(1)=1$. It follows that $T_\nu \circ\glog\in\gL$.
In order to find sufficient conditions for $T_\nu \circ\glog$ to have no such pole, 
we first look at the case $\nu>0$. In that case no pole exists if 
$\underline{\glog}=\min\{\glog(x)|x\in\mathbb{R}^+\}\geq -1/\nu$. In other words,  
$\nu\leq-1/\underline{\glog}=\nu_+[\glog]$.  
Now we turn to the case $\nu<0$. Then no pole exists if $\overline{\glog}=\max\{\glog(x)|x\in\mathbb{R}^+\}\leq -1/\nu$, which is to say 
$\nu\geq-1/\overline{\glog}=\nu_-[\glog]$.  
Both cases together show that $T_\nu \circ\glog$ 
is a continuous function only if 
$\nu_-[\glog]\leq \nu\leq \nu_+[\glog]$. 
Finally, if $T_\nu \circ\glog$ is continuous, 
then $\max\{T_\nu\circ\glog(x)|x\in\mathbb{R}^+\}=T_\nu(\max\{\glog(x)|x\in\mathbb{R}^+\})$, i.e. 
$\overline{T_\nu\circ\glog}=T_\nu(\overline{\glog})$. Analogously we find $\underline{T_\nu\circ\glog}=T_\nu(\underline{\glog})$.
From this 
$\nu_-[T_\nu\circ\glog]=-1/T_\nu(\overline{\glog})=-1/\overline{\glog}-\nu=\nu_-[\glog]-\nu$ follows.
An analogous relation holds for $\nu_+$, which completes the proof.         
\linebreak
\linebreak
%
\linebreak
{\bf Theorem 2} 
Suppose a map $\mystar$ is given on $\gL$ with the properties (i) $*\circ *$ is the identity map, (ii) for each $\glog\in\gL$ there exists a $\nu^*$ such that $\glog^*=T_{\nu^*} \circ \glog$, and (iii) $*$ commutes with scale transformations $g\in\gG$ (that is $(\glog\circ g)^\mystar=\glog^\mystar\circ g$). Then $\mystar$ is uniquely determined and $\nu^*$
is given by
\begin{equation}
\nu^{\mystar}=\nu_++\nu_-\,,\quad\mbox{for}\quad\glog\in\gL_{\nu_+,\nu_-}\, .
\label{theor2equ}
\end{equation}
Furthermore it follows from theorem (1) that
\begin{equation}
\glog\in\gL_{\nu_+,\nu_-}\Leftrightarrow \, \glog^{\mystar}\in\gL_{-\nu_-,-\nu_+}\,.
\label{dualsets}
\end{equation}
\linebreak
{\bf Proof of Theorem 2.}
The duality $\mystar$ on $\gL$ can be constructed in the following way. From properties (ii) stated in theorem (2) we know
there exists a 
functional $F:\gL\to \mathbb{R}$ 
such that $\glog^\mystar=T_{F[\glog]}\circ \glog$. From property (i) and (ii) 
we also know that $\glog=T_{F[\glog^\mystar]}\circ \glog^\mystar$.
Theorem (1) states that given $\glog\in\gL_{\nu_+,\nu_-}$ the condition 
$\nu_+\geq F[\glog]\geq \nu_-$ is necessary 
for $\glog^*$ to be a proper logarithm. As a consequence we get 
$T_{F[\glog]}\circ \glog\in\gL_{\nu_+-F[\glog],\nu_--F[\glog]}$. This further
implies $\nu_+-F[\glog]\geq F[\glog^\mystar]\geq \nu_--F[\glog]$.
Property (iii)
implies that for any two logarithms $\glog_1$ and $\glog_2$ that are members 
of the same class $\gL_{\nu_+,\nu_-}$, we get 
$F[\glog_1]=F[\glog_2]$.
Therefore $F$ can only be of the form
\begin{equation}
	F[\glog]=f(\nu_+,-\nu_-)\quad\mbox{for}\quad \glog\in\gL_{\nu_+,\nu_-}\,,
\end{equation}
where $f:\mathbb{R}_+^2\to\mathbb{R}$. 
Using $T_\nu\circ T_\mu=T_{\nu+\mu}$ 
together with 
property (i)
leads to
\begin{equation}
	f(\nu_+,-\nu_-)=-f(\nu_+-f(\nu_+,-\nu_-),-(\nu_--f(\nu_+,-\nu_-))) \,.
\end{equation}
In other words, $f$ solves the two equations
\begin{equation}
\begin{array}{lc}
	\mbox{(a)}\,\,& f(x,y)=-f(x-f(x,y),y+f(x,y))\,
	\\ & \\
	\mbox{(b)}\,\,& x\geq f(x,y)\geq -y\,
	\end{array}
	\label{A1}
\end{equation}
for all $x,y\in\mathbb{R}_+$.
Eq. (\ref{A1} b) immediately tells us that $f(0,0)=0$.
%
Consider a function 
$y(x,z)$ solving the implicit equation
\begin{equation}
f(x,y(x,z))=z\, ,
\label{A2}
\end{equation}
and rewrite Eq. (\ref{A1} a) 
as
\begin{equation}
f(x-z,y(x,z)+z)=-z\,.
\label{A3}
\end{equation}
From Eqs. (\ref{A2}) and (\ref{A3}) one gets
$y(x-z,-z)=y(x,z)+z$.
Also, $f(0,0)=0$ implies $y(0,0)=0$.
By expanding $y(x,z)=\sum_{m,n=0}^{\infty}y_{m,n}x^m z^n$
one gets  $y_{0,0}=0$ and for the first order
\begin{equation}
2y_{0,1}+y_{1,0}+1=0\,.
\label{A4}
\end{equation}
All coefficients $y_{m,n}$ of higher order follow equation 
\begin{equation}
y_{m,n}=(-1)^n\sum_{k=0}^n{m+k\choose{m}}y_{m+k,n-k}\,.
\end{equation}
We also expand $f(x,y(x,z))=\sum_{m,n=1}^{\infty} f_{m,n}x^m y(x,z)^n=\sum_{i,j=1}^{\infty} \phi_{i,j}x^iz^j=z$. 
It follows that all $\phi_{i,j}=0$ except for $\phi_{0,1}=1$. Since $y_{0,0}=0$ all terms $f_{m,n}x^my(x,z)^n$ 
can only contribute to coefficients $\phi_{i,j}$ with indices $i\geq n$ or $j\geq n$. Comparing 
coefficients order by order one shows that only coefficients of the first order
\begin{equation}
f_{1,0}x+f_{0,1}y(x,z)=z\,,
\end{equation}
contribute to solving Eq. (\ref{A2}).
Thus $y(x,z)$ can only be of the form $y(x,z)=y_{1,0}x+y_{0,1}z$.
This, together with Eqs. (\ref{A4}) and (\ref{A1} b), implies
$x\geq 2(y_{1,0}x-y)/(1+y_{1,0}) \geq -y$. 
Choosing $x=0$ 
gives $1\geq y_{1,0}\geq -1$. Setting $y=0$ also implies $y_{1,0}\geq 1$,
so that the only possible solution for $y_{1,0}$ is 
$y_{1,0}=1$. As a consequence of Eq. (\ref{A4}), $y_{0,1}=-1$. Therefore we have 
$y(x,z)=x-z$ and $f$ has the unique solution $f(x,y)=x-y$.
From this it follows that 
$\nu^{\mystar}=f(\nu_+,-\nu_-)=\nu_++\nu_-$ for $\glog\in\gL_{\nu_+,\nu_-}$.
This means that $\nu^{\mystar}$ is uniquely defined.
Since $\nu_+-\nu^{\mystar}=-\nu_-$ and $\nu_--\nu^{\mystar}=-\nu_+$, theorem (1) 
implies that $\glog\in\gL_{\nu_+,\nu_-}\Leftrightarrow \, \glog^{\mystar}\in\gL_{-\nu_-,-\nu_+}$.
This completes the proof.

\end{article}
\end{document}